\documentclass[prd,onecolumn,nofootinbib]{revtex4}
\usepackage{bm,amsmath,amssymb}

\begin{document}

\title{Gravity with spin excludes fermionic strings}
\author{Nikodem Pop{\l}awski}
\affiliation{Department of Physics, Indiana University, Swain Hall West, 727 East Third Street, Bloomington, IN 47405, USA}
\date{\today}

\begin{abstract}
The existence of intrinsic spin of matter requires the metric-affine formulation of gravity, in which the affine connection is not constrained to be symmetric and its antisymmetric part (torsion tensor) is a dynamical variable.
We show that the cyclic identity for the curvature tensor in the metric-affine formulation forbids fermions represented by Dirac spinors to form point or string configurations.
Consequently, fermionic strings contradict the gravitational field equations in the presence of spin.
Superstring theory is therefore incorrect.
\end{abstract}

\maketitle

\section{Angular momentum without gravitational field}

We consider a physical system in the absence of the gravitational field, described by a matter Lagrangian density $\mathfrak{L}$ which depends on matter fields $\phi$, their first derivatives $\phi_{,i}$ with respect to the coordinates $x^i$, and $x^i$ \cite{Niko,KS}.
Under an infinitesimal coordinate transformation $x^i\rightarrow x^{'i}=x^i+\xi^i$, where $\xi^i=\delta x^i$ is a variation of $x^i$, the Lagrangian density $\mathfrak{L}$ transforms like a scalar density: $\delta \mathfrak{L}=(|\partial x^i/\partial x^{'i}|-1)\mathfrak{L}=-\xi^i_{\phantom{i},i}\mathfrak{L}$.
The variation $\delta\mathfrak{L}$ is also equal to $\delta\mathfrak{L}=(\partial\mathfrak{L}/\partial\phi)\delta\phi+\bigl(\partial\mathfrak{L}/\partial(\phi_{,i})\bigr)\delta(\phi_{,i})+(\bar{\partial}\mathfrak{L}/\partial x^i)\xi^i$, where $\bar{\partial}$ denotes partial differentiation with respect to $x^i$ at constant $\phi$ and $\phi_{,i}$.
Using the Lagrange equations $\partial\mathfrak{L}/\partial\phi-\partial_i\bigl(\partial\mathfrak{L}/\partial(\phi_{,i})\bigr)=0$, and the identities $\mathfrak{L}_{,i}=\bar{\partial}\mathfrak{L}/\partial x^i+(\partial\mathfrak{L}/\partial\phi)\phi_{,i}+\bigl(\partial\mathfrak{L}/\partial(\phi_{,j})\bigr)\phi_{,ji}$ and $\delta(\phi_{,i})=(\delta\phi)_{,i}-\xi^j_{\phantom{j},i}\phi_{,j}$, leads to the conservation law:
\begin{equation}
\mathfrak{J}^i_{\phantom{i},i}=0,
\label{Noeth6}
\end{equation}
for the current vector density
\begin{equation}
\mathfrak{J}^i=\xi^i\mathfrak{L}+\frac{\partial\mathfrak{L}}{\partial(\phi_{,i})}(\delta\phi-\xi^j\phi_{,j}).
\label{Noeth7}
\end{equation}
The existence of a conservation law for each continuous symmetry of a Lagrangian density formulates the Noether theorem.

For Lorentz rotations, $\xi^i=\epsilon^i_{\phantom{i}j}x^j$ and $\delta\phi=(1/2)\epsilon_{ij}G^{ij}\phi$, where $\epsilon_{ij}=-\epsilon_{ji}$ are infinitesimal quantities and $G^{ij}$ are the generators of the Lorentz group.
The corresponding current (\ref{Noeth7}) is
\begin{equation}
\mathfrak{J}^i=\epsilon^{kl}\biggl(\frac{\partial\mathfrak{L}}{\partial(\phi_{,i})}\phi_{,[l}x_{k]}-\delta^i_{[l}x_{k]}\mathfrak{L}+\frac{1}{2}\frac{\partial\mathfrak{L}}{\partial(\phi_{,i})}G_{kl}\phi\biggr),
\end{equation}
where $[\,]$ denotes antisymmetrization.
Because $\epsilon^{kl}$ are arbitrary, (\ref{Noeth6}) gives the conservation law:
\begin{equation}
\mathfrak{M}_{kl\phantom{i},i}^{\phantom{kl}i}=0,
\label{clLg8}
\end{equation}
for the angular momentum density
\begin{equation}
\mathfrak{M}_{kl}^{\phantom{kl}i}=x_k\theta^{\phantom{l}i}_l-x_l\theta^{\phantom{k}i}_k+\frac{\partial\mathfrak{L}}{\partial(\phi_{,i})}G_{kl}\phi,
\label{clLg9}
\end{equation}
where 
\begin{equation}
\theta^{\phantom{i}k}_i=\frac{\partial\mathfrak{L}}{\partial(\phi_{,k})}\phi_{,i}-\delta^k_i \mathfrak{L}
\label{cemd2}
\end{equation}
is the canonical energy-momentum density.
The first two terms on the right-hand side of (\ref{clLg9}) form the orbital angular momentum density, and the last term is the canonical spin density $\Sigma_{kl}^{\phantom{kl}i}$.
The conservation law (\ref{clLg8}) for the angular momentum density can be written as the conservation law for the canonical spin density in the special theory of relativity \cite{Niko,KS}:
\begin{equation}
\Sigma_{kl\phantom{i},i}^{\phantom{kl}i}=\theta_{kl}-\theta_{lk}.
\label{clLg12}
\end{equation}

\section{Angular momentum with gravitational field}

In the metric-affine formulation of gravity, the tetrad $e^i_a$ and the spin connection $\omega^a_{\phantom{a}bk}=e^a_j e^j_{\phantom{j}b;k}=e^a_j(e^j_{\phantom{j}b,k}+\Gamma^{\,\,j}_{i\,k}e^i_b)$ are the dynamical variables describing the geometry of spacetime \cite{Niko,KS,Hehl,rev}.
Semicolon denotes the covariant derivative with respect to the affine connection $\Gamma^{\,\,i}_{j\,k}$.
The affine connection is asymmetric in the lower indices and its antisymmetric part is the torsion tensor \cite{Niko,KS,Hehl,rev}:
\begin{equation}
S^i_{\phantom{i}jk}=\Gamma^{\,\,\,\,i}_{[j\,k]}.
\end{equation}
The spin connection appears in the covariant derivative of a Lorentz vector: $V^a_{\phantom{a}|i}=V^a_{\phantom{a},i}+\omega^a_{\phantom{a}bi}V^b$, analogously to $\Gamma^{\,\,i}_{j\,k}$ in the covariant derivative of a vector, $V^k_{\phantom{k};i}=V^k_{\phantom{k},i}+\Gamma^{\,\,k}_{l\,i}V^l$.
The tetrad relates spacetime coordinates $i,j,...$ to local Lorentz coordinates $a,b,...$: $V^a=V^i e^a_i$.
Its covariant derivative vanishes: $e^a_{i|k}=0$, where vertical bar denotes the covariant derivative acting on both spacetime and Lorentz coordinates.
Lorentz coordinates are thus lowered or raised by the Minkowski metric tensor $\eta_{ab}$ of a flat spacetime, analogously to the metric tensor $g_{ik}$ lowering or raising spacetime coordinates.
The metricity condition $g_{ij;k}=0$ gives the affine connection $\Gamma^{\,\,k}_{i\,j}=\{^{\,\,k}_{i\,j}\}+C^k_{\phantom{k}ij}$, where $\{^{\,\,k}_{i\,j}\}=(1/2)g^{km}(g_{mi,j}+g_{mj,i}-g_{ij,m})$ are the Christoffel symbols, $C^i_{\phantom{i}jk}=S^i_{\phantom{i}jk}+2S_{(jk)}^{\phantom{(jk)}i}$ is the contortion tensor, and $(\,)$ denotes symmetrization.
It also constrains the spin connection to be antisymmetric in its Lorentz indices: $\omega^{ab}_{\phantom{ab}i}=-\omega^{ba}_{\phantom{ba}i}$.
Instead of $e^i_a$ and $\omega^{ab}_{\phantom{ab}i}$, the metric tensor $g_{ik}$ and the torsion tensor $S^i_{\phantom{i}jk}$ can be taken as the dynamical variables.

We consider a physical system in the presence of the gravitational field.
The variation of $\mathfrak{L}_\textrm{m}$ with respect to the spin connection defines the dynamical spin density \cite{Niko,KS,Hehl,rev}
\begin{equation}
\mathfrak{S}_{ab}^{\phantom{ab}i}=2\frac{\delta\mathfrak{L}_\textrm{m}}{\delta\omega^{ab}_{\phantom{ab}i}}=2\frac{\partial\mathfrak{L}_\textrm{m}}{\partial\omega^{ab}_{\phantom{ab}i}},
\label{spden1}
\end{equation}
which is antisymmetric in the Lorentz indices: $\mathfrak{S}_{ab}^{\phantom{ab}i}=-\mathfrak{S}_{ba}^{\phantom{ba}i}$.
The spin tensor is defined as
\begin{equation}
s_{ijk}=\frac{1}{\mathfrak{e}}\mathfrak{S}_{ijk},
\label{tensor}
\end{equation}
where $\mathfrak{e}=\mbox{det}\,e^a_i=\sqrt{-\mbox{det}\,g_{ik}}$.
The second equality in (\ref{spden1}) is satisfied because a matter Lagrangian density may depend on the spin connection but not on its derivatives; a scalar density depending on derivatives of $\omega^{ab}_{\phantom{ab}i}$ is a Lagrangian density for the gravitational field.
The spin density is also given by $\mathfrak{S}_{ij}^{\phantom{ij}k}=2\delta\mathfrak{L}_\textrm{m}/\delta C^{ij}_{\phantom{ij}k}$.
The variation of the Lagrangian density for matter $\mathfrak{L}_\textrm{m}$ with respect to the tetrad defines the dynamical energy-momentum density \cite{Niko,KS,Hehl,rev}
\begin{equation}
\mathfrak{T}^{\phantom{i}a}_i=\frac{\delta\mathfrak{L}_\textrm{m}}{\delta e^i_a}.
\label{dtemd2}
\end{equation}
The metric energy-momentum tensor $T_{ij}=(2/\mathfrak{e})(\delta\mathfrak{L}_\textrm{m}/\delta g^{ij})$ is related to the dynamical energy-momentum density and the spin tensor by the Belinfante-Rosenfeld relation: $T_{ik}=\mathfrak{T}_{ik}/\mathfrak{e}-(1/2)(s_{ik}^{\phantom{ik}j}-s_{k\phantom{j}i}^{\phantom{k}j}+s^j_{\phantom{j}ik})_{;j}+S_j(s_{ik}^{\phantom{ik}j}-s_{k\phantom{j}i}^{\phantom{k}j}+s^j_{\phantom{j}ik})$.
Since the variations $\delta\omega^{ab}_{\phantom{ab}i}$ are independent of $\delta e^i_a$, the spin density is independent of the energy-momentum density.

The Lorentz group is the group of tetrad rotations, $e^a_i=\Lambda^a_{\phantom{a}b}e^b_i$, where $\Lambda^a_{\phantom{a}b}$ is a Lorentz matrix.
Since a matter Lagrangian density $\mathfrak{L}_\textrm{m}(\phi,\phi_{,i})$ is invariant under local, proper Lorentz transformations, it is invariant under tetrad rotations: $\delta\mathfrak{L}_\textrm{m}=(\partial\mathfrak{L}_\textrm{m}/\partial\phi)\delta\phi+\bigl(\partial\mathfrak{L}_\textrm{m}/\partial(\phi_{,i})\bigr)\delta(\phi_{,i})+\mathfrak{T}^{\phantom{i}a}_i\delta e^i_a+(1/2)\mathfrak{S}_{ab}^{\phantom{ab}i}\delta\omega^{ab}_{\phantom{ab}i}=0$, where the changes $\delta$ are caused by a tetrad rotation.
Upon integration of $\delta\mathfrak{L}_\textrm{m}$ over spacetime, the terms with $\phi$ and $\phi_{,i}$ vanish because of the Lagrange equations:
\begin{equation}
\int\Bigl(\mathfrak{T}^{\phantom{i}a}_i\delta e^i_a+\frac{1}{2}\mathfrak{S}_{ab}^{\phantom{ab}i}\delta\omega^{ab}_{\phantom{ab}i}\Bigr)d\Omega=0.
\label{cam2}
\end{equation}
For an infinitesimal Lorentz transformation, $\Lambda^a_{\phantom{a}b}=\delta^a_b+\epsilon^a_{\phantom{a}b}$, where $\epsilon^a_{\phantom{a}b}=-\epsilon_b^{\phantom{b}a}$ are infinitesimal quantities, the tetrad $e^a_i$ changes by $\delta e^a_i=\Lambda^a_{\phantom{a}b}e^b_i-e^a_i=\epsilon^a_{\phantom{a}i}$, and the tetrad $e_a^i$ changes by $\delta e_a^i=-\epsilon_{\phantom{i}a}^i$.
Accordingly, the spin connection changes by $\delta\omega^{ab}_{\phantom{ab}i}=\delta(e^a_j \omega^{j b}_{\phantom{j b}i})=\epsilon^a_{\phantom{a}j}\omega^{j b}_{\phantom{j b}i}-e^a_j \epsilon^{j b}_{\phantom{j b};i}=\epsilon^a_{\phantom{a}c}\omega^{cb}_{\phantom{cb}i}-e^a_j \epsilon^{j b}_{\phantom{j b}|i}+\epsilon^a_{\phantom{a}c}\omega^{bc}_{\phantom{bc}i}=-\epsilon^{ab}_{\phantom{ab}|i}$.
Substituting these variations into (\ref{cam2}) and using partial integration $\int\mathfrak{V}^i_{\phantom{i};i}d\Omega=2\int S_i \mathfrak{V}^i d\Omega$, where $\mathfrak{V}$ is any contravariant vector density and $S_i=S^k_{\phantom{k}ik}$, leads to $-\int\Bigl(\mathfrak{T}^{\phantom{i}a}_i\epsilon^i_{\phantom{i}a}+\frac{1}{2}\mathfrak{S}_{ab}^{\phantom{ab}i}\epsilon^{ab}_{\phantom{ab}|i}\Bigr)d\Omega=-\int\Bigl(\mathfrak{T}_{ij}\epsilon^{ij}+\frac{1}{2}\mathfrak{S}_{ij}^{\phantom{ij}k}\epsilon^{ij}_{\phantom{ij}|k}\Bigr)d\Omega=\int\Bigl(-\mathfrak{T}_{[ij]}-S_k\mathfrak{S}_{ij}^{\phantom{ij}k}+\frac{1}{2}\mathfrak{S}_{ij\phantom{k};k}^{\phantom{ij}k}\Bigr)\epsilon^{ij}d\Omega=0$.
Since the infinitesimal Lorentz rotation $\epsilon^{ij}$ is arbitrary, we obtain the conservation law for the spin density \cite{Niko,KS,Hehl,rev}:
\begin{equation}
\mathfrak{S}_{ij\phantom{k};k}^{\phantom{ij}k}-2S_k\mathfrak{S}_{ij}^{\phantom{ij}k}=\mathfrak{T}_{ij}-\mathfrak{T}_{ji}.
\label{cam7}
\end{equation}
This law can be written as
\begin{equation}
\mathfrak{S}^{ijk}_{\phantom{ijk},k}-\Gamma^{\,\,i}_{l\,k}\mathfrak{S}^{jlk}+\Gamma^{\,\,j}_{l\,k}\mathfrak{S}^{ilk}-2\mathfrak{T}^{[ij]}=0.
\label{Pap22}
\end{equation}

A matter Lagrangian density $\mathfrak{L}_\textrm{m}$ can be written as $\mathfrak{L}_\textrm{m}=\mathfrak{e}L$, where $L$ is a scalar.
If $\mathfrak{L}_\textrm{m}$ depends on matter fields $\phi$ (minimally coupled to the affine connection) and their first derivatives $\phi_{,i}$, and the fields $\phi$ do not contain vector indices, then the tetrad appears in $L$ only through derivatives of $\phi$, in a covariant combination $e^i_a \phi_{|i}$.
Such fields can be, for example, spinor fields.
Varying $\mathfrak{L}_\textrm{m}$ with respect to the tetrad gives $\delta\mathfrak{L}_\textrm{m}=\mathfrak{e}\delta L-\mathfrak{e}e^a_i L\delta e^i_a=\mathfrak{e}\bigl(\partial L/\partial(\phi_{|a})\bigr)\phi_{|i}\delta e^i_a-\mathfrak{L}_\textrm{m}e^a_i \delta e^i_a=\Bigl(\bigl(\partial\mathfrak{L}_\textrm{m}/\partial(\phi_{|a})\bigr)\phi_{|i}-e^a_i\mathfrak{L}_\textrm{m}\Bigr)\delta e^i_a$.
The dynamical energy-momentum density (\ref{dtemd2}) is therefore $\mathfrak{T}^{\phantom{i}a}_i=\bigl(\partial\mathfrak{L}_\textrm{m}/\partial(\phi_{|a})\bigr)\phi_{|i}-e^a_i \mathfrak{L}_\textrm{m}$.
The corresponding tensor with two coordinate indices,
\begin{equation}
\mathfrak{T}^{\phantom{i}k}_i=\frac{\partial\mathfrak{L}_\textrm{m}}{\partial(\phi_{|k})}\phi_{|i}-\delta^k_i \mathfrak{L}_\textrm{m}=\frac{\partial\mathfrak{L}_\textrm{m}}{\partial(\phi_{,k})}\phi_{|i}-\delta^k_i \mathfrak{L}_\textrm{m},
\label{cemd3}
\end{equation}
generalizes the canonical energy-momentum density (\ref{cemd2}) to the presence of the gravitational field \cite{Niko,rev}.
The spin connection $\omega^{ab}_{\phantom{ab}i}$ appears in $\mathfrak{L}_\textrm{m}$ only through derivatives of $\phi$, in a combination $-\bigl(\partial\mathfrak{L}_\textrm{m}/\partial(\phi_{,i})\bigr)\Gamma_i\phi$, where $\Gamma_i=-(1/2)\omega_{abi}G^{ab}$ is the connection in the covariant derivative of $\phi$: $\phi_{|i}=\phi_{,i}-\Gamma_i\phi$.
The dynamical spin density (\ref{spden1}) is therefore $\mathfrak{S}_{ab}^{\phantom{ab}i}=\bigl(\partial\mathfrak{L}_\textrm{m}/\partial(\phi_{,i})\bigr)G_{ab}\phi$.
The corresponding tensor with two coordinate indices, $\mathfrak{S}_{kl}^{\phantom{kl}i}=\bigl(\partial\mathfrak{L}_\textrm{m}/\partial(\phi_{,i})\bigr)G_{kl}\phi$, coincides with the canonical spin density $\Sigma_{kl}^{\phantom{kl}i}$ in (\ref{clLg9}).
Consequently, the conservation law (\ref{cam7}) for the spin density generalizes (\ref{clLg12}) to the presence of the gravitational field \cite{Niko,KS}.

In the metric formulation of gravity, the tetrad (or the metric tensor) is the only dynamical variable representing the gravitational field \cite{LL}.
In that formulation, (\ref{cam2}) reduces to $\int\mathfrak{T}^{\phantom{i}a}_i\delta e^i_a d\Omega=0$ and (\ref{cam7}) reduces to $\mathfrak{T}_{ij}=\mathfrak{T}_{ji}$, which is not a generalization of (\ref{clLg12}).
The torsion tensor is constrained to vanish and the affine connection is equal to the Levi-Civita connection (Christoffel symbols), so the spin connection is a function of the tetrad and its first derivatives.
Accordingly, the variations $\delta\omega^{ab}_{\phantom{ab}i}$ are functions of $\delta e^i_a$ and their derivatives.
The density (\ref{spden1}) is thus a function of the energy-momentum density, forming a part of the orbital angular momentum density.
The metric formulation therefore excludes intrinsic spin of matter.
Consequently, the existence of intrinsic spin requires the metric-affine formulation and the torsion tensor.

\section{Dirac spinors in spacetime with torsion}

Elementary particles, that are fermions, are described by Dirac spinor fields (spinor wave functions).
In the metric-affine formulation of gravity, the Dirac Lagrangian density for a free spinor $\psi$ with mass $m$, minimally coupled to the gravitational field, is given by $\mathfrak{L}_\textrm{m}=(i/2)\hbar c\mathfrak{e}(\bar{\psi}\gamma^k\psi_{;k}-\bar{\psi}_{;k}\gamma^k\psi)-mc^2\mathfrak{e}\bar{\psi}\psi$, where $\bar{\psi}=\psi^{\dag}\gamma^0$ is the adjoint spinor corresponding to $\psi$ \cite{Niko,rev}.
The covariant derivative of $\psi$, $\psi_{;k}=\psi_{,k}-\Gamma_k\psi$, gives $\bar{\psi}_{;k}=\bar{\psi}_{,k}+\bar{\psi}\Gamma_k$.
The Dirac matrices $\gamma^a$ obey $\gamma^{(a}\gamma^{b)}=\eta^{ab}I$ and transform under local Lorentz transformations like $\psi\bar{\psi}$.
The last relation yields $\gamma^a_{\phantom{a}|k}=\omega^{a}_{\phantom{a}bk}\gamma^b-[\Gamma_k,\gamma^a]$, which gives the Fock-Ivanenko spinor connection $\Gamma_k=-(1/4)\omega_{abk}\gamma^a\gamma^b$, in accordance with the generators of the Lorentz group in the spinor representation, $G^{ab}=(1/2)\gamma^{[a}\gamma^{b]}$.
Varying the total action for the gravitational field and fermionic matter with respect to the adjoint spinor $\bar{\psi}$ and equaling this variation to zero gives the Dirac equation $i\hbar\gamma^k\psi_{;k}=mc\psi$.

The energy-momentum tensor corresponding to the Dirac Lagrangian is equal to $T_{ik}=(i/2)\hbar c(\bar{\psi}\delta^j_{(i}\gamma_{k)}\psi_{;j}-\bar{\psi}_{;j}\delta^j_{(i}\gamma_{k)}\psi)-(i/2)\hbar c(\bar{\psi}\gamma^j\psi_{;j}-\bar{\psi}_{;j}\gamma^j\psi)g_{ik}+mc^2\bar{\psi}\psi g_{ik}$.
The covariant derivative of a spinor can be decomposed into the covariant derivative with respect to the Levi-Civita connection and a term containing the contortion tensor: $\psi_{;k}=\psi_{:k}+(1/4)C_{ijk}\gamma^{[i}\gamma^{j]}\psi$, $\bar{\psi}_{;k}=\bar{\psi}_{:k}-(1/4)C_{ijk}\bar{\psi}\gamma^{[i}\gamma^{j]}$.
The contortion tensor therefore appears in the Dirac Lagrangian density in a term $(i/8)\hbar c\bar{\psi}(\gamma^k\gamma^{[i}\gamma^{j]}+\gamma^{[i}\gamma^{j]}\gamma^k)\psi C_{ijk}$.
Consequently, the spin tensor for a Dirac field is completely antisymmetric \cite{Niko,KS,Hehl,rev}:
\begin{equation}
s^{ijk}=s^{[ijk]}=-\frac{1}{\mathfrak{e}}\epsilon^{ijkl}s_l,\,\,\,s^i=\frac{1}{2}\hbar c\bar{\psi}\gamma^i\gamma^5\psi,
\label{spin}
\end{equation}
where $\epsilon^{ijkl}$ is the Levi-Civita permutation symbol, $\gamma^5=i\gamma^0\gamma^1\gamma^2\gamma^3$, and $s^i$ is the Dirac spin pseudovector.
The spin tensor (\ref{spin}) does not depend on $m$, and remains the same if we include the electromagnetic, weak or strong interactions of fermions.
Since the covariant derivative of a spinor appears only in the kinetic term in the Lagrangian density, $(i/2)\hbar c\mathfrak{e}(\bar{\psi}\gamma^k\psi_{;k}-\bar{\psi}_{;k}\gamma^k\psi)$, and the Lagrangian density is additive, the spin density is also additive.
Accordingly, the spin tensor for a system of fermions is also completely antisymmetric.

\section{Einstein-Cartan-Sciama-Kibble theory}

The Lagrangian density for the gravitational field contains the first derivatives of the spin or affine connection, which appear through the curvature tensor, $R^a_{\phantom{a}bij}=\omega^a_{\phantom{a}bj,i}-\omega^a_{\phantom{a}bi,j}+\omega^a_{\phantom{a}ci}\omega^c_{\phantom{c}bj}-\omega^a_{\phantom{a}cj}\omega^c_{\phantom{c}bi}$ or $R^i_{\phantom{i}mjk}=\partial_{j}\Gamma^{\,\,i}_{m\,k}-\partial_{k}\Gamma^{\,\,i}_{m\,j}+\Gamma^{\,\,i}_{l\,j}\Gamma^{\,\,l}_{m\,k}-\Gamma^{\,\,i}_{l\,k}\Gamma^{\,\,l}_{m\,j}$ \cite{Niko,KS,rev}.
This tensor satisfies the Bianchi identity, $R^i_{\phantom{i}n[jk;l]}=2R^i_{\phantom{i}nm[j}S^m_{\phantom{m}kl]}$, and the cyclic identity, $R^m_{\phantom{m}[jkl]}=-2S^m_{\phantom{m}[jk;l]}+4S^m_{\phantom{m}n[j}S^n_{\phantom{n}kl]}$ \cite{Niko,rev,Scho}.
The curvature tensor can be decomposed as $R^i_{\phantom{i}klm}=P^i_{\phantom{i}klm}+C^i_{\phantom{i}km:l}-C^i_{\phantom{i}kl:m}+C^j_{\phantom{j}km}C^i_{\phantom{i}jl}-C^j_{\phantom{j}kl}C^i_{\phantom{i}jm}$, where $P^i_{\phantom{i}klm}$ is the Riemann tensor and colon denotes the covariant derivative with respect to the Levi-Civita connection.
The Ricci tensor is given by $R^a_{\phantom{a}i}=R^{ab}_{\phantom{ab}ij}e^j_b$ or $R_{ik}=R^j_{\phantom{j}ijk}$.

The simplest and most natural gravitational Lagrangian density is a linear function of the curvature tensor, given by
\begin{equation}
\mathfrak{L}_\textrm{g}=-\frac{1}{2\kappa}\mathfrak{e}R,
\label{ecsk}
\end{equation}
where $R=R^b_{\phantom{b}j}e^j_b=R^i_{\phantom{i}i}$ is the Ricci scalar and $\kappa=8\pi G/c^4$ is Einstein's gravitational constant (which sets the units of mass).
Such a function has no free parameters.
Varying the total action for the gravitational field and matter, $S=(1/c)\int(\mathfrak{L}_\textrm{g}+\mathfrak{L}_\textrm{m})d\Omega$, with respect to the torsion tensor (or the spin connection) and equaling this variation to zero gives the Cartan field equations \cite{Niko,KS,Hehl,rev}
\begin{equation}
S^j_{\phantom{j}ik}-S_i \delta^j_k+S_k \delta^j_i=-\frac{1}{2}\kappa s^{\phantom{ik}j}_{ik}.
\label{Cartan}
\end{equation}
These equations are linear and algebraic: torsion is proportional to the density of intrinsic spin of matter and thus vanishes outside material bodies.
Varying the total action $S$ with respect to the tetrad and equaling this variation to zero gives the Einstein field equations \cite{Niko,KS,Hehl,rev}
\begin{equation}
R_{ki}-\frac{1}{2}Rg_{ik}=\frac{\kappa}{\mathfrak{e}}\mathfrak{T}_{ik}.
\label{Einstein}
\end{equation}
Contracting the Bianchi identity and substituting into it (\ref{Cartan}) and (\ref{Einstein}) gives the conservation law for the dynamical energy-momentum density: $\mathfrak{T}^{ij}_{\phantom{ij}:j}=C_{jk}^{\phantom{jk}i}\mathfrak{T}^{jk}+(1/2)\mathfrak{S}_{klj}R^{klji}$.
Contracting the cyclic identity and substituting into it (\ref{Cartan}) and (\ref{Einstein}) yields the conservation law (\ref{cam7}) for the spin density.
A more complicated Lagrangian density for the gravitational field would give more complicated field equations.
Those equations, however, upon substituting into the contracted Bianchi and cyclic identities would still give the same conservation laws.
The existence of intrinsic spin is therefore related to the cyclic identity for the curvature tensor.

Varying the total action $S$ with respect to the metric tensor and equaling this variation to zero gives the Riemannian form of the Einstein equations, $G_{ik}=\kappa(T_{ik}+U_{ik})$, where $G_{ik}$ is the Einstein tensor and
\begin{equation}
U^{ik}=\kappa\biggl(-s^{ij}_{\phantom{ij}[l}s^{kl}_{\phantom{kl}j]}-\frac{1}{2}s^{ijl}s^k_{\phantom{k}jl}+\frac{1}{4}s^{jli}s_{jl}^{\phantom{jl}k}+\frac{1}{8}g^{ik}(-4s^l_{\phantom{l}j[m}s^{jm}_{\phantom{jm}l]}+s^{jlm}s_{jlm})\biggr)
\label{Ein}
\end{equation}
is a contribution to the energy-momentum tensor from torsion, which is quadratic in the spin tensor \cite{rev}.
The spin tensor also appears in $T_{ik}$ because $\mathfrak{L}_\textrm{m}$ depends on torsion.
The metric-affine formulation of gravity, based on the Lagrangian density (\ref{ecsk}), constitutes the Einstein-Cartan-Sciama-Kibble (ECSK) theory of gravity \cite{Niko,KS,Hehl,rev}, and the corresponding metric formulation constitutes the general theory of relativity (GR) \cite{LL}.
Since the metric-affine formulation of gravity is required by the existence of intrinsic spin, the ECSK theory is more complete than GR.
The corrections from the spin tensor to the right-hand side of the Einstein equations are significant only at extremely high densities, above the Cartan density, for which the square of the density of spin is on the order of the energy density multiplied by $\kappa$.
Below the Cartan density, the predictions of the ECSK theory do not differ from the predictions of GR.
In vacuum, where torsion vanishes, this theory reduces to GR.
The ECSK gravity therefore passes all observational and experimental tests of GR \cite{exp}.

Substituting the completely antisymmetric spin tensor for a Dirac field (\ref{spin}) into the Cartan equations (\ref{Cartan}) gives the completely antisymmetric torsion tensor:
\begin{equation}
S_{ijk}=C_{ijk}=\frac{1}{2}\kappa e_{ijkl}s^l.
\label{torsion}
\end{equation}
Consequently, the Dirac equation can be written as a nonlinear (cubic) equation for $\psi$ \cite{HD}:
\begin{equation}
i\hbar\gamma^k\psi_{:k}=mc\psi-\frac{3}{8}\hbar^2 c\kappa(\bar{\psi}\gamma^k\gamma^5\psi)\gamma_k\gamma^5\psi.
\label{Dirac}
\end{equation}
Its adjoint conjugate is $-i\hbar\bar{\psi}_{:k}\gamma^k=mc\bar{\psi}-(3/8)\hbar^2 c\kappa(\bar{\psi}\gamma^k\gamma^5\psi)\bar{\psi}\gamma_k\gamma^5$.
Substituting (\ref{spin}) into the Einstein equations (\ref{Einstein}) gives $U^{ik}=(1/4)\kappa(2s^i s^k+s^l s_l g^{ik})$.
Putting the Dirac equation in the energy-momentum tensor for a Dirac field gives $T_{ik}=(i/2)\hbar c(\bar{\psi}\delta^j_{(i}\gamma_{k)}\psi_{;j}-\bar{\psi}_{;j}\delta^j_{(i}\gamma_{k)}\psi)=(i/2)\hbar c(\bar{\psi}\delta^j_{(i}\gamma_{k)}\psi_{:j}-\bar{\psi}_{:j}\delta^j_{(i}\gamma_{k)}\psi)+(1/2)\kappa(-s_i s_k+s^l s_l g_{ik})$.
The corresponding combined energy-momentum tensor, which appears in the Riemannian form of the Einstein equations, is thus \cite{cosm}
\begin{equation}
T_{ik}+U_{ik}=\frac{i}{2}\hbar c(\bar{\psi}\delta^j_{(i}\gamma_{k)}\psi_{:j}-\bar{\psi}_{:j}\delta^j_{(i}\gamma_{k)}\psi)+\frac{3}{4}\kappa s^l s_l g_{ik}.
\label{combined}
\end{equation}
The first term on the right-hand side of (\ref{combined}) is the GR part of the energy-momentum tensor for a Dirac field.
It can be macroscopically averaged at cosmological scales as a perfect fluid.
The averaged second term gives in the comoving frame of reference the spinor-torsion contributions to the energy density of the fluid, $-(9/16)\kappa(\hbar cn)^2$, and its pressure, $(9/16)\kappa(\hbar cn)^2$, where $n$ is the fermion number density.
These contributions are significant in the early Universe, where they avert the unphysical big-bang singularity (predicted by GR), replacing it with a cusp-like bounce at a finite minimum scale factor, before which the Universe was contracting \cite{cosm}.
They also explain why the observable Universe at largest scales appears spatially flat, homogeneous and isotropic, without needing cosmic inflation \cite{infl}.
The simplest metric-affine formulation of gravity, based on the Lagrangian density (\ref{ecsk}), therefore not only includes the intrinsic spin of matter in the geometry of spacetime, but also naturally removes several cosmological problems that exist in the metric formulation.

\section{Mutltipole expansion of energy-momentum and spin densities}

We consider matter which is distributed over a small region in space and consists of points with coordinates $x^i$, forming an extended body whose motion is represented by a world tube in spacetime \cite{Pap,NSH}.
The motion of the body as a whole is represented by an arbitrary timelike world line $C$ inside the world tube, which consists of points with coordinates $X^i(s)$, where $s$ is the affine parameter along $C$.
We assume that the dimensions of the body are small, so the tensors $\mathfrak{T}$ and $\mathfrak{S}$ describing the extended body are different from zero only within a sphere whose center is the point with coordinates $X^i$ and whose radius $R$ is very small for any time $t$.
When $R\rightarrow0$, the arbitrariness in the choice of $C$ disappears.
We define
\begin{equation}
\delta x^i=x^i-X^i,\,\,\,u^i=\frac{dX^i}{ds}.
\label{Pap1}
\end{equation}
We also define the integrals \cite{NSH}:
\begin{eqnarray}
& & M^{ik}=u^0\int\mathfrak{T}^{ik}dV, \label{Pap2} \\
& & M^{ijk}=-u^0\int\delta x^i\mathfrak{T}^{jk}dV, \label{Pap3} \\
& & N^{ijk}=u^0\int\mathfrak{S}^{ijk}dV, \label{Pap4}
\end{eqnarray}
where the integration is carried over the volume hypersurface (three-dimensional space) for $t=$ const.
Accordingly, $\delta x^0=0$ and $M^{0jk}=0$.
Higher multipole integrals contain more factors $\delta x^i$.
The $\alpha$-component of spin is proportional to $\epsilon_{\alpha ijk}N^{ijk}$.

In the dipole approximation, integrals with two or more factors $\delta x^i$ multiplying $\mathfrak{T}^{jk}$ and integrals with one or more factors $\delta x^i$ multiplying $\mathfrak{S}^{jkl}$ can be neglected.
Integrating (\ref{Pap22}) over the volume hypersurface and using Gau\ss-Stokes theorem to eliminate surface integrals gives $\int\mathfrak{S}^{ij0}_{\phantom{ij0},0}dV-\int\Gamma^{\,\,i}_{l\,k}\mathfrak{S}^{jlk}dV+\int\Gamma^{\,\,j}_{l\,k}\mathfrak{S}^{ilk}dV-2\int\mathfrak{T}^{[ij]}dV=0$.
The conservation law (\ref{Pap22}) also gives $(x^l\mathfrak{S}^{ijk})_{,k}=\mathfrak{S}^{ijl}+x^l\Gamma^{\,\,i}_{l\,k}\mathfrak{S}^{jlk}-x^l\Gamma^{\,\,j}_{l\,k}\mathfrak{S}^{ilk}+2x^l\mathfrak{T}^{[ij]}$, which upon integrating over the volume hypersurface and eliminating surface integrals leads to $\int(x^l\mathfrak{S}^{ij0})_{,0}dV=\int\mathfrak{S}^{ijl}dV+\int x^l\Gamma^{\,\,i}_{m\,k}\mathfrak{S}^{jmk}dV-\int x^l\Gamma^{\,\,j}_{m\,k}\mathfrak{S}^{imk}dV+2\int x^l\mathfrak{T}^{[ij]}dV$.
The last relation upon substituting (\ref{Pap1}) becomes $(u^l/u^0)\int\mathfrak{S}^{ij0}dV+X^l\int\mathfrak{S}^{ij0}_{\phantom{ij0},0}dV=\int\mathfrak{S}^{ijl}dV+2\int\delta x^l\mathfrak{T}^{[ij]}dV+X^l(\int\Gamma^{\,\,i}_{m\,k}\mathfrak{S}^{jmk}dV-\int\Gamma^{\,\,j}_{m\,k}\mathfrak{S}^{imk}dV+2\int\mathfrak{T}^{[ij]}dV)$.
The integrated (\ref{Pap22}) reduces it to $(u^l/u^0)\int\mathfrak{S}^{ij0}dV=\int\mathfrak{S}^{ijl}dV+2\int\delta x^l\mathfrak{T}^{[ij]}dV$, which is equivalent to \cite{NSH}
\begin{equation}
M^{l[ij]}=-\frac{1}{2}\biggl(\frac{u^l}{u^0}N^{ij0}-N^{ijl}\biggr).
\label{Pap29}
\end{equation}
Putting $l=0$ in (\ref{Pap29}) gives the identity.
For a Dirac field, the complete antisymmetry of the spin density (\ref{spin}) gives
\begin{equation}
N^{ijk}=N^{[ijk]}.
\label{antisym}
\end{equation}

\section{Fermions cannot be point particles}

Let us assume that a fermionic field forms a zero-dimensional (point) configuration.
For such a configuration located at the origin of Cartesian coordinates, the energy-momentum density $\mathfrak{T}^{ik}$ at radius vector ${\bf r}$ is proportional to $\delta({\bf r})$:
\begin{equation}
\mathfrak{T}^{ik}({\bf r})=v^{ik}\delta({\bf r}),
\label{point}
\end{equation}
where $v^{ik}$ is a finite quantity describing the fermion as a whole.
Accordingly, $M^{ik}$ is finite: $M^{ik}=u^0 v^{ik}$.
We also have $M^{\alpha ij}=-u^0 v^{ij}\int\int\int\delta x^\alpha \delta(x)\delta(y)\delta(z)dx\,dy\,dz$, where
\begin{equation}
\delta x^\alpha=x^\alpha
\end{equation}
and $\alpha$ denotes spatial coordinates, which leads to
\begin{equation}
M^{\alpha ij}=0.
\end{equation}
Consequently, (\ref{Pap29}) reduces to
\begin{equation}
N^{ijl}=\frac{u^l}{u^0}N^{ij0}.
\label{eq1}
\end{equation}
For a Dirac spinor field $\psi$, which satisfies (\ref{antisym}), putting $j=0$ in (\ref{eq1}) gives
\begin{equation}
N^{il0}=-\frac{u^l}{u^0}N^{i00}=0.
\label{eq2}
\end{equation}
Substituting (\ref{eq2}) into (\ref{eq1}) gives \cite{NSH,non}
\begin{equation}
N^{ijl}=0.
\label{eq3}
\end{equation}

The spin density $\mathfrak{S}^{ijk}$ at radius vector ${\bf r}$ is proportional to $\delta({\bf r})$: $\mathfrak{S}^{ijk}({\bf r})=v^{ijk}\delta({\bf r})$, where $v^{ijk}$ is a finite quantity describing the fermion as a whole.
Accordingly, $N^{ijl}$ is finite: $N^{ijl}=u^0 v^{ijl}$.
The relation (\ref{eq3}) yields $v^{ijl}=0$ and thus
\begin{equation}
\mathfrak{S}^{ijl}=0.
\label{none}
\end{equation}
Hence, the conservation law (\ref{cam7}) yields the symmetry of the energy-momentum tensor, $\mathfrak{T}^{ij}=\mathfrak{T}^{ji}$.
The relation (\ref{none}), through (\ref{tensor}) and (\ref{spin}), can be true only if $\psi=0$: in the absence of the spinor.
Equivalently, if $\psi\neq0$ then point distributions (\ref{point}) cannot represent fermions, otherwise the conservation law (\ref{cam7}) for the spin density and thus the invariance of the matter Lagrangian density under local Lorentz transformations (tetrad rotations) would be violated.
In the metric-affine formulation of gravity, the relation (\ref{Pap29}) follows from the cyclic identity for the curvature tensor.
Consequently, the point approximation of a Dirac field is not a solution of the gravitational field equations in the metric-affine formulation \cite{non}.

Since a Dirac field cannot form a point distribution, fermions cannot be point particles \cite{non}.
This conclusion seems natural because higher moments in the multipole expansion should be included to encode angular momentum, hence spin.
Moreover, a Dirac field cannot be a system of points because each point has a symmetric energy-momentum tensor, for which (\ref{Pap3}) would give $M^{i[jk]}=0$.
In this case, (\ref{Pap29}) and (\ref{antisym}) would again lead to (\ref{none}) and $\psi=0$.
Because torsion in the metric-affine formulation of gravity prevents Dirac fields from forming point configurations, it also determines the minimal spatial extension $d$ of a fermion represented by such a field.
The size of this extension is given by a condition at which torsion introduces significant corrections to the energy-momentum tensor.
In the ECSK theory, such corrections are significant when $U_{ik}$ (\ref{Ein}) is on the order of $T_{ik}$.
Equivalently, this size is determined by a condition at which the cubic term in the Dirac equation (\ref{Dirac}) is on the order of the mass term.
The energy-momentum tensor for a Dirac field is on the order of $mc^2|\psi|^2$ (in the rest frame of the fermion), the spin tensor is on the order of $\hbar c|\psi|^2$, and the wave function $\psi\sim d^{-3/2}$.
The spatial extension $d$ is thus on the order of the Cartan length $l_C$ defined by \cite{non}
\begin{equation}
\frac{m}{l_C^3}=\frac{G}{c^4}\biggl(\frac{\hbar}{l_C^3}\biggr)^2.
\label{order}
\end{equation}
The left-hand side of (\ref{order}) defines the Cartan density $\rho_C$.

For an electron, $l_C\sim10^{-27}$ m is 15 orders of magnitude smaller than its Compton wavelength and 8 orders of magnitude greater than the Planck length.
Quantum gravity effects, which may be significant at the Planck length, therefore do not affect the prediction of the ECSK theory that fermions are spatially extended on the order of $l_C$.
A discovery of physical dimension of fermions around their Cartan lengths would confirm the ECSK theory.
Spatial extension of fermions described by Dirac wave fuctions indicates that fermionic point configurations are not allowed if the Dirac field is quantized.
Accordingly, the two-point function of a nonlinear spinor theory based on (\ref{Dirac}) should exhibit self-regulation of its short-distance behavior \cite{Mer}.
Torsion may thus introduce a natural ultraviolet cutoff in quantum field theory at distances on the order of $l_C$ \cite{non}.

\section{Fermions cannot be strings}

Let us assume that a fermionic field forms a one-dimensional (string) configuration of finite length.
In cylindrical coordinates, $(x^1=r,x^2=\phi,x^3=z)$, such a curve can be represented by two functions, $r_0(\phi)$ and $z_0(\phi)$.
These functions 
The energy-momentum density $\mathfrak{T}^{ik}$ at radius vector ${\bf r}$ is proportional to $\delta(r-r_0)\delta(z-z_0)$:
\begin{equation}
\mathfrak{T}^{ik}({\bf r})=v^{ik}\delta(r-r_0(\phi))\delta(z-z_0(\phi)),
\label{string}
\end{equation}
where $v^{ik}$ is a finite quantity describing the fermion as a whole.
Accordingly, $M^{ik}$ is finite: $M^{ik}=u^0 v^{ik}\int r_0(\phi)d\phi$.
We also have $M^{\alpha ij}=-u^0 v^{ij}\int\int\int\delta x^\alpha\delta(r-r_0(\phi))\delta(z-z_0(\phi))rdr\,d\phi\,dz$, where
\begin{equation}
\delta x^1=r-r_0(\phi),\,\,\,\delta x^3=z-z_0(\phi),
\end{equation}
which leads to
\begin{equation}
M^{1ij}=M^{3ij}=0.
\end{equation}
Consequently, (\ref{Pap29}) gives
\begin{equation}
N^{ij1}=\frac{u^1}{u^0}N^{ij0},\,\,\,N^{ij3}=\frac{u^3}{u^0}N^{ij0}.
\label{eq4}
\end{equation}
For a Dirac spinor field $\psi$, which satisfies (\ref{antisym}), putting $j=0$ in (\ref{eq4}) gives
\begin{equation}
N^{i10}=-\frac{u^1}{u^0}N^{i00}=0,\,\,\,N^{i30}=-\frac{u^3}{u^0}N^{i00}=0.
\label{eq5}
\end{equation}
Thus $N^{012}=N^{013}=N^{023}=0$.
Substituting (\ref{eq5}) into (\ref{eq4}) gives $N^{123}=0$.
Therefore, string distributions (\ref{string}) must also obey the relation (\ref{eq3}).

The spin density $\mathfrak{S}^{ijk}$ at radius vector ${\bf r}$ is proportional to $\delta(r-r_0)\delta(z-z_0)$: $\mathfrak{S}^{ijk}({\bf r})=v^{ijk}\delta(r-r_0)\delta(z-z_0)$, where $v^{ijk}$ is a finite quantity describing the fermion as a whole.
Accordingly, we obtain $N^{ijl}=u^0 v^{ijl}\int r_0(\phi)d\phi$.
Because $\int r_0(\phi)d\phi>0$, the relation (\ref{eq3}) yields $v^{ijl}=0$ and thus (\ref{none}), which can be true only if $\psi=0$: in the absence of the spinor.
This result remains valid if $r_0(\phi)$ and $z_0(\phi)$ are multivalued.
Also, this result does not depend on whether the string is open or closed.
Furthermore, the $\alpha$-component of spin is proportional to $\epsilon^{\alpha jkl}N_{jkl}$, so (\ref{eq3}) indicates that the spin of a string configuration given by (\ref{string}) vanishes.
The spin of a fermion, however, is different from zero.
Consequently, string distributions (\ref{string}) cannot represent fermions, otherwise the conservation law (\ref{cam7}) for the spin density and thus the invariance of the matter Lagrangian density under local Lorentz transformations (tetrad rotations) would be violated.
Because the relation (\ref{Pap29}) follows from the cyclic identity for the curvature tensor, the classical string description of Dirac spinors is not a solution of the gravitational field equations in the metric-affine formulation of gravity.
Therefore, quantization of fermionic strings is also inconsistent with metric-affine gravity.

Superstring theory, or supersymmetric string theory, assumes that all elementary fermions are oscillating, one-dimensional strings whose length is on the order of the Planck length \cite{POl}.
Although that theory lacks predictive power \cite{anti}, it has become the most popular contender for a theory of everything \cite{pro}.
The results of this paper show that one-dimensional fermionic strings contradict the cyclic identity for the curvature tensor in the metric-affine formulation of gravity which is required in the presence of intrinsic spin.
In addition, the simplest (ECSK) gravity with spin requires fermions to be spatially extended on the order of their Cartan lengths which are much greater than the Planck length.
Consequently, fermionic strings contradict two combined physical phenomena that are well-established by experiment and observation: the geometric nature of the gravitational field and the existence of spin.
Fermionic strings are thus unphysical and superstring theory is incorrect.

\end{document}